\documentclass{aa}
\usepackage{graphicx}
\usepackage{txfonts}

\begin{document}

\title{An evolutionary catalogue of galactic post-AGB and related objects}
\subtitle{}
\author{R. Szczerba\inst{1}
           \and N. Si\'odmiak\inst{2}
           \and G. Stasi\'nska\inst{3}
           \and J. Borkowski\inst{1}
               }

          \offprints{R. Szczerba, \\ \email{szczerba@ncac.torun.pl}}

          \institute{N. Copernicus Astronomical Center, Rabia{\'n}ska 8,
                    87-100 Toru{\'n}, Poland
                \and
                    Space Telescope Science Institute, 3700 San Martin
Drive, Baltimore, MD 21218, USA
                 \and
                    LUTH, Observatoire de Meudon, 5 Place Jules Janssen,
                    F-92195 Meudon Cedex, France
                    }

          \date{Received / Accepted }

             \abstract
{}
{With the ongoing AKARI infrared sky survey, of much greater
sensitivity than IRAS, a wealth of post-AGB
objects may be discovered. It is thus time to organize our present
knowledge of known post-AGB stars in
the galaxy with a view to using it to search for new post-AGB objects
among AKARI sources.}
{We searched  the literature available on the NASA Astrophysics
Data System up to 1 October 2006, and defined criteria for classifying
sources into three categories: \emph{very likely}, \emph{possible} and
\emph{disqualified} post-AGB objects.  The category of \emph{very likely}
post-AGB objects is made up of several classes.}
{We have created an evolutionary, on-line catalogue of
Galactic post-AGB objects, to be referred to as \emph{the Toru\'n
catalogue of Galactic post-AGB and related objects}. The present
version of the catalogue contains 326 \emph{very likely}, 107 
\emph{possible} and
64 \emph{disqualified}
objects. For the very likely post-AGB objects, the catalogue gives the
available optical and infrared photometry,
infrared spectroscopy and spectral types, and links to finding charts
and bibliography.}
{}


\titlerunning{A catalogue of Galactic post-AGB objects }
\authorrunning{ R. Szczerba et al.}

\maketitle

\section{Introduction}

Once they have burnt up their central hydrogen and helium low- and 
intermediate-mass
ascend the asymptotic giant branch (AGB), where they
burn hydrogen and helium in thin shells surrounding the core. This
phase, during
which the atmosphere expands and cools down, is accompanied by
intense mass loss
($10^{-7} - 10^{-4}$\,M$_\odot$\,yr$^{-1}$). When the mass of the
remaining H--rich envelope becomes
smaller than $\sim$ $10^{-2}$\,M$_\odot$ (the exact value is very much a
function of the initial
mass, see, for example, Sch{\"o}nberner\,\cite{S83}), the envelope begins to
shrink and the effective
temperature  starts to increase (Paczy\'nski\,\cite{P71}). If the
temperature increases
on a timescale shorter than the dispersion time of the matter
previously ejected
by the star, a planetary nebula (PN) will appear, as the result of
the ionization
of the circumstellar shell and of the emission of optical line radiation. The
intermediate phase between the end of the AGB phase and the PN phase
is called the
post-AGB phase. It was formerly also referred to as the proto-PN phase or the
pre-PN phase (see, for example, Lo \& Bechis\,\cite{LB76}, Likkel et
al.\,\cite{LMO87}, Hrivnak,
Kwok \& Volk\,\cite{HKV89}, Geballe et al.\,\cite{GTK92},
Kwok\,\cite{K93}). We prefer
the term post-AGB object\footnote{The term post-AGB star has also 
sometimes been used
to denote very hot objects, e.g. PG\,1159 stars, which are different
objects from the ones considered in this paper and in our 
catalogue.}, since this also covers objects that have
left the AGB stage
but will evolve to the white dwarf stage without ever becoming a
planetary nebula.
The first review devoted to such objects was by
Zuckerman\,(\cite{Z78}). Important material on post-AGB objects can
be found in the
Conference Proceedings edited by Kwok \& Pottasch\,(\cite{KP87}),
Preite-Martinez\,(\cite{PM87}), Schwarz\,(\cite{S93}), Le Bertre,
L$\grave{\rm e}$bre, \& Waelkens\,(\cite{LBLW99}),
Szczerba \& G{\'o}rny\,(\cite{SG01}), and in the book by Habing \&
Olofsson\,(\cite{HO04}),
as well as in the review papers by Kwok\,(\cite{K93}) and van
Winckel\,(\cite{VW03}).

Post-AGB stars are difficult -- and often impossible -- to detect in
optical wave lenghts,
being strongly obscured by their circumstellar dusty envelopes. Infrared sky
surveys such as the 2 $\mu$m Sky Survey (Neugebauer \&
Leighton\,\cite{NL69}), the Air
Force Infrared Sky Survey (Kleinmann et al.\,\cite{KGJ81}) and
especially the InfraRed Astronomical
Satellite (IRAS) survey (Beichman et al.\,\cite{BNH88}) were
essential for the discoveries
of post-AGB objects in the past. With the ongoing  AKARI
(Shibai\,\cite{S04}) infrared sky survey,
of much greater sensitivity than IRAS, a wealth of post-AGB objects
will be discovered. It is thus
time to organize our present knowledge of post-AGB stars in our
Galaxy into an evolutionary,
on-line catalogue that can easily be updated by future discoveries.

A preliminary compilation for this catalogue was presented by
Szczerba et al.\,(\cite{SGZJ01}) and included
about 220 objects. We have now scanned all the literature
available on the NASA Astrophysics Data System (ADS) up to 1 October,
2006.  In its present
version, the catalogue, to be referred to as `'the Toru{\'n}
catalogue of Galactic post-AGB and related
objects'',  contains 326 {\it very likely}, 107 \emph{possible} and 64
{\it disqualified} post-AGB objects.
A subsample of the {\it very likely} objects  was used in the work by
Stasi\'nska et al.\,(\cite{SSS06}), which showed that post-AGB stars,
as a class, are
valuable testbeds for theories of AGB nucleosynthesis, and in the
work by Si\'odmiak et al.(\cite{SMU07}),
devoted to the Hubble Space Telescope snapshot survey of post-AGB
objects characterized by different physical
and chemical properties.

\section{Classification criteria}

Post-AGB sources  form a very heterogenous group of stars. A most difficult
task in creating this catalogue was to define which criteria  the
objects should satisfy in order to be said to be post-AGB stars. 
In general, the
criteria are based on the central star properties (spectral type and luminosity
class) and/or circumstellar envelope (disc?) properties (infrared
excess, spectral
dust and/or gas features). However, for many sources data exists only for one
of those components. In some cases the star is not
directly observable and the only available information is
about the envelope. Such a situation is expected for
higher mass sources where the timescales of stellar evolution are
much shorter than the
timescales of envelope dispersion into the interstellar medium. In
other cases, there is some
information on the central star (luminosity class I--III and
appropriate spectral type) but no
(or only a small)  infrared (IR) excess is detected. Such a situation
is  typical for lower mass
sources with very long timescales of stellar evolution.

In defining criteria, we have been inspired by the precepts
introduced by Kwok\,(\cite{K93}),
Van Winckel\,(\cite{VW03}) and Waelkens \& Waters\,(\cite{WW04}). We
emphasize that with the
established criteria we have tried to cover all stellar objects
classified for various reasons as
post-AGB\footnote{The exception are non-variable OH/IR stars -- see below}.
The coolest post-AGB objects included in our catalogue have
spectral types not later than
K, while the hottest have effective temperatures below 25\,000\,K
(see Szczerba et al.\,\cite{SGZJ01}). We count as {\it disqualified}
post-AGB objects those that have central stars classified as M- or
as cool AGB carbon stars.
We have also disqualified luminosity class V objects, as well as
objects that have been classified as planetary nebulae (especially
those appearing in the Strasbourg-ESO Catalogue of Galactic
Planetary Nebulae  -- Acker\,\cite{AMO92}). Post-AGB objects
collected in our database include the following classes (which are
not mutually exclusive):

\begin{itemize}

\item {\bf IRAS selected sources} -- {\it IRASsel} in the catalogue.
These sources were considered to be
post-AGB on
the  basis of their IRAS colors, found to be  between those of AGB
stars and those of planetary nebulae
(e.g. Kwok et al.\,\cite{KBH87}; van der Veen et al.\,\cite{vdVHG89};
Hu et al.\,\cite{HSdJ93},
Garcia-Lario\,\cite{GLMP97}  or, most recently, Su{$\acute{\rm
a}$}rez et al.\,\cite{SGLM06}).
Most of the objects selected in this way are optically faint since
the selection criteria focus on the
circumstellar material properties. Note, however, that there is a
significant overlap between IRAS color-selected, post-AGB sources and
other classes of objects, like
galaxies, AGB stars, planetary nebulae or
young stellar objects (see for example Fig.3 in  Su{$\acute{\rm a}$}rez et
al.\,\cite{SGLM06}).
Therefore, additional observations are usually necessary to classify them
properly.

\item {\bf High Galactic latitude supergiants} -- {\it hglsg} in the catalogue.
Massive supergiants are generally not expected to be found at high
Galactic latitudes. It is therefore likely that high-latitude 
supergiants are in fact low-mass stars in their final
stages of evolution.  Bidelman\,(\cite{B51}) was the first to point
out the existence of
such supergiants (e.g. \object{89\,Her}). However, some of the high-latitude
supergiants could also be
massive Population I stars that have been ejected from the Galactic
disk. Subsequent analyses of chemical
abundances confirmed that very often the high-latitude
supergiants are of low metallicity
(in agreement with the interpretation that they are Population II supergiants).
The spectral types of such sources are found to be K, G, F and A.  In addition,
some members of this class of post-AGB objects show extreme Fe-group
element depletion (up to -4 dex
or even more). Since the abundance of zinc is only mildly weak, the
abundance of Fe-group
elements cannot reflect the primordial value and a selective removal of metals
from the photosphere through grain formation and mass loss has been
proposed for those objects
(van Winckel et al.\cite{VWMW92}).
We consider all the {\it very likely } post-AGB objects which have 
$|b|>$15{\degr} to belong to the {\it hglsg} class  (except
those classified as {\it hglB} -- see below).

\item {\bf High Galactic latitude B-type supergiants} -- {\it hglB}
in the catalogue and
{\bf hot post-AGB objects} -- {\it hotpAGB}. Hot, B-type supergiants were
discovered in studies of the B-star population in the Galactic halo.
Since they are
similar to  main sequence B-type stars (effective temperature,
gravity), detailed chemical
composition studies are required to classify  these objects properly
(e.g. McCausland et al.\,\cite{MCCD92}, Moehler \& Heber\,\cite{MH98}
and references therein).
We classify as {\it hglB} such {\it very likely} post-AGB objects
(appropriate chemical composition),
which have B spectral type and are located at high Galactic latitudes
($|b|>$15{\degr}). There
exists also a group of early-type (A and B)
supergiants, located at lower latitudes ($|b|<$15{\degr}), which have
been classified as post-AGB on the
basis of their chemical composition  (see e.g. Parthasarathy et
al.\,\cite{PGF01}, Moehler\,\cite{M01},
Parthasarathy\,\cite{P04} and references therein). Such sources were
included in our catalogue as
{\it hotpAGB}. Note that some of those sources (the more massive ones
?) still display IR excess (e.g. Conlon et al.\,\cite{CDK93}).

\item {\bf Bright stars with infrared (IR) excess} -- {\it IRexc} in
the catalogue.
The IR excess criterion proposed by Parthasarathy \& Pottasch\,(\cite{PP86}),
Pottasch \& Parthasarathy\,(\cite{PP88}) and others, together with
dust mass estimation, is one
of the criteria that have been used to select candidate post-AGB sources.
However, not all the sources selected in this way can be considered
to be post-AGB objects.
For example, from the list of 10  infrared excess sources suggested by
Pottasch \& Parthasarathy\,(\cite{PP88}) to be post-AGB objects, we
classified five as {\it very likely}, one as a \emph{possible} (we 
cannot decide if
the object is Herbig Ae/Be or post-AGB), and three 
as {\it disqualified} post-AGB
objects (Herbig Ae/Be stars). For
one of their objects (\object{SAO\,253680}) there is probably a mistake 
in terms of its identification in the SIMBAD
database. Among the eight objects considered by Hrivnak et
al.\,(\cite{HKV89}), on the basis of their infrared excess, as \emph{possible}
post-AGB objects,  six appear to be {\it very likely}
post-AGB and two are {\it disqualified}
in our catalogue. From the list of 21 post-AGB objects (including RV
Tau stars) in Oudmaijer et
al.\,(\cite{OvdVW92}), there are 18 which remain as {\it very likely}
post-AGB objects in our catalogue
while three are {\it disqualified}.

\item {\bf UU Her-type stars} -- {\it UU Her} in the catalogue. Among
Population II
supergiants, there is a small group of variable stars called UU Her
after the best
studied member of this group\footnote{Note, however, that Klochkova et
al.\,(\cite{KPC97}), after a  detailed chemical composition analysis of UU Her
concluded that this star is a low-ass supergiant but not a post-AGB
star. Therefore,
\object{UU\,Her} is included in our catalogue as a {\it disqualified} post-AGB
object, while other members of the
group  are still classified as {\it UU Her}.}
(Sasselov\,\cite{S84}, Bartkevi$\check{\rm c}$ius\,\cite{B92}). The
characteristic
criteria of UU Her-type stars, in addition to the general {\it hglsg}
properties,
include high radial velocities, small amplitude pulsations
(of order of 0.1 mag) and large IR excess due to circumstellar dust.
Bartkevi$\check{\rm c}$ius\,(\cite{B92}) collected the available
information on 18
UU Her-type stars, and we classified  12 of them as {\it very likely}
post-AGB, three as {\it possible} post-AGB
and  three are {\it disqualified} in our catalogue. The 
\object{PS\,Gem} is not in the Bartkevi$\check{\rm c}$ius\,(\cite{B92}) list,
but the object is classified in the literature as UU Her-type as well.

\item {\bf RV Tau stars} - {\it RV Tau} in the catalogue. The RV Tau
stars are highly luminous variable
stars, which show alternating deep and shallow minima, periods between
30 and 150 days,
and spectral types F, G and K (e.g. Preston et al.\,\cite{PKS63}).
Most of them show
IR excess, which has been interpreted by Jura\,(\cite{J86}) as
evidence of a recent
mass loss typical of the AGB phase of
evolution. All RV Tauri stars with near-IR excess in their 
energy distribution
(the dusty ones)
seem to be binaries, while this is not the case for the naked ones 
(e.g. Van Winckel et al.\,2000).
We have considered all the sources classified
as RV Tau stars in the SIMBAD database. However, in a few cases we
found that they had to be {\it disqualified} in the light of the most recent 
research (e.g.\object{R\,Sct} see Matsura et al.\,\cite{MYZ02}).

\item {\bf R CrB stars} -- {\it R CrB}; {\bf extreme helium stars} --
{\it eHe}; and
{\bf Late Thermal Pulse objects} --
{\it LTP} in the catalogue. The R CrB stars are rare H-deficient and
C-rich supergiants that undergo
irregular declines of up to 8 magnitudes when dust forms in clumps along the
line of sight (see e.g. Clayton\,\cite{C96} for a review). The
extreme H-deficiency of
the R CrB stars suggests that some mechanism removed the entire H-rich stellar
envelope. There are two major models which explain their origin: a merger
scenario (Webbink\,\cite{W84}, Iben \& Tutukov\,\cite{IT85}) and a
final helium-shell flash
scenario (Fujimoto\,\cite{F77}, Renzini\,\cite{R79}, Iben et al.\cite{IKT83}).
There is still no consensus on
which of the two scenarios is valid (none of them can explain all the observed
properties), but only the second scenario implies a post-AGB nature.
We have  included 36 R CrB stars as {\it very likely} post-AGB objects in our
catalogue, but it could be that they are not bona-fide post-AGB stars
but simply related objects. Extreme helium stars (16 known
with effective temperature below 25\,000\,K),
which could be evolutionarily connected to R CrB stars (see e.g. Pandey
et al.\,\cite{PKL01}) are also
included in our catalogue. In addition we have included two {\it
LTP} stars (\object{Sakurai's\,object} and \object{FG\,Sge}).
On the other hand, we did  not include \object{V605\,Aql},
which has experienced a late thermal pulse, but has presently an
effective temperature larger than 50\,000\,K.

\item {\bf 21 micron emission sources} -- {\it 21 micron} in the
catalogue. There is a group of 12 sources
which show a spectacular emission band at 21 microns (Kwok et
al.\,\cite{KVH89}) confirmed by
the Infrared Space Observatory (ISO) observations. All these sources
are C-rich and show s-process
element enhancement (e.g. Van Winckel \& Reyniers\,\cite{VWR00}).
Since this dust
feature is still not well understood we have included information on
this class of objects in our catalogue.

\item {\bf Reflection nebulosity} -- {\it refneb} in the catalogue.
Finally, for a few famous objects
(\object{Red\,Rectangle}, \object{Minkowski\,Footprint}, \object{Egg\,Nebula},
\object{AFGL\,618}) we have indicated the characteristic which
allowed these sources to be discovered prior to IRAS.

\end{itemize}

There is one group of objects
considered as post-AGB which was intentionally left out of the
present edition of the catalogue. These are non-variable OH/IR stars
(Habing et al.\,\cite{HvdVG87}). Molecular emission in the
post-AGB phase of evolution is so important and there are so many
interesting results that we plan to
cover this topic in an upcoming  edition of the catalogue.
Similarly, the problem of nebular
morphology is not covered in the present edition and has been left to
the future development of this
evolutionary catalogue.

In the light of the above criteria, we have drawn up three lists. The
first of these - {\it very likely} post-AGB
objects - contains 326 sources. An object can belong to this list only
if it satisfies at least one of
the criteria above and if there is a significant number of references
in ADS (at least 5) that do not
contradict its post-AGB classification. Using this somewhat arbitrary
criterion we considered as
{\it possible} post-AGB objects those sources that satisfy one of
the above criteria but have less
than five papers referring to them in ADS. This criterion does not apply
in the case of {\it RV Tau},
{\it R CrB} and {\it eHe}, as there is a consensus on the nature of
these objects. We have
also classified as {\it possible} post-AGB those objects for which
contradictory opinions appear in
the literature and it is not clear which of them is correct. For such
cases we have indicated why we
cannot treat the object as a {\it very likely}
post-AGB and have listed the corresponding references. Altogether, we
have 107 {\it possible}
post-AGB sources in our catalogue. The third part of the catalogue
contains {\it disqualified} objects,
i.e.  objects that were once considered in the literature as post-AGB
objects or candidates, but for
which subsequent observations or different interpretations of the
data showed that they can no longer be
considered as such. This list contains 64 objects. For each of them,
we indicate the reason for their disqualification.

The distribution of the  {\it very likely} post-AGB objects into
various classes is shown in Table\,\ref{pAGBclass}. Note that one
object may belong to several classes.

\begin{table}
      \caption[]{The distribution of the {\it very likely} post- AGB
objects into various
classes.}
      \label{pAGBclass}
      \centering
      \begin{tabular}{l r}
      \hline\hline
      Class      &  number of objects \\
      \hline
      {\it IRASsel} & 115  \\
      {\it hglsgl} & 63  \\
      {\it hglB} & 9  \\
      {\it hotpAGB} & 18  \\
      {\it IRexc} & 34  \\
      {\it UU Her} & 13  \\
      {\it RV Tau} & 99 \\
      {\it R CrB} & 36 \\
      {\it eHe} & 16 \\
      {\it LTP} & 2 \\
      {\it 21 micron} & 12 \\
      {\it refneb} & 4 \\
      \hline
      \end{tabular}
     \end{table}

There are 218 objects classified as post-AGB in the SIMBAD catalogue.
Of these only 107 are among our {\it very likely} post-AGB objects, 53 are
{\it possible} post-AGB, 35 have been  {\it disqualified}, IRAS\,05298$-$6957
is a LMC object, and we are still keeping 22 objects on the ``waiting list''
of OH-selected sources as possible non-variable OH/IR stars, which are
not considered in this edition of the catalogue. There are 99 stars
classified as RV Tau in the SIMBAD
database. Of these 80 are in our list of {\it very likely} post-AGB
objects, two are {\it possible}, 12 are {\it disqualified}
and five are not considered in the catalogue since they are LMC sources.
On the other hand, we included in the catalogue
19 RV Tauri stars, which are not classified as RV Tau in the SIMBAD
database. They are mostly from the recent work of
De Ruyter\,(\cite{DRVWM06}). There are  48 sources classified as R
CrB-type stars in the SIMBAD database. Of these 33 belong to our list
of {\it very likely} Galactic post-AGB sources, two are in the SMC and 13 in
the LMC. In our catalogue we have, in addition, three sources that are
not counted as R CrB by SIMBAD (\object{V3795\,Sgr}, \object{V348\,Sgr} and
\object{UX\,Ant}).
On the other hand, our catalogue contains 66 sources (not classified as
{\it RV Tau}, {\it R CrB}, {\it eHe} or {\it LTP} objects)  which are
not considered as post-AGB in the SIMBAD database.

Recently, De Ruyter et
al.\,(\cite{DRVWM06})  presented a systematic study of the
spectral energy distribution  of a
sample of 51 post-AGB objects and listed the confirmed binary
systems. The photospheres of several
binary systems also feature the characteristic depletion pattern
discussed in the group of
{\it hglsg} objects. Waters et al.\,(\cite{WTW92}) argued that the
most favorable
conditions for efficient dust-gas separation occur if the
circumstellar dust is trapped in a stable
disk configuration. Binary evolution may be responsible for the
observed morphology changes
in post-AGB objects and in planetary nebulae. (For a review see
Sahai\,\cite{S01} and Balick
\& Frank\,\cite{BF02}). It is worthwhile pointing out that binarity has a 
non-negligible role
in attracting our attention to such objects, which would have been 
much less conspicuous
otherwise. Therefore, available information about confirmed binarity
of post-AGB objects is included in our catalogue together with the
corresponding reference listed in the ``Bibliography'' field  (see
below). Altogether we have found arguments for the binarity of 47 ``very 
likely'' post-AGB
objects.

\section{Description of the catalogue}

The on-line version of the catalogue (available at
http://www.ncac.torun.pl/postagb) is organized into
3 sub-catalogues. On the entry screen  the user can choose between
``very likely post-AGB",
``possible post-AGB", and ``disqualified objects". When selecting one
of the above buttons, the screen indicates the total number of
objects of this category in our database and lists them in order of
increasing Galactic longitude
and latitude ($l$,\,$b$)  (column 1), the IRAS identification (column
2), HD and SAO
identifications (column 3 and 4, respectively), other names, if
any\footnote{``other name" is either the
usual name or the designation in one of the following main catalogues
(in order of preference):
General Catalogue of Variable Stars;  LS or LSE; BD; CD catalogues
and any other name deemed appropriate
(if none of the above main catalogues has info about the object).}
(column 5), classification(-s)
(column 6), information about binarity and spectral type\footnote{For
the vast majority of sources the
spectral type is taken from the SIMBAD database. Some exceptions
are marked by the {\it bulb} icon
in our catalogue.  Keeping the mouse positioned on the {\it bulb}
icon for 2-3 seconds makes a yellow box pop up 
with more detailed information about the bibliographical source
from which we derived the spectral type.}
(column 7 and 8, respectively), 
and the bibliography related to the object (the ADS link
gives connection to ADS with
the full listing of bibliography, while the papers we selected  can
be reached by clicking on the
corresponding acronyms). The full list of the selected bibliographic
entries can be seen by clicking
``Bibliography" on the main bar above the object list. As a rule we
did not add selected bibliographic
entries when the total number of references for a given object in ADS
was less than 10, except for the newest references: De Ruyter et
al.\,(\cite{DRVWM06}) and Su{$\acute{\rm a}$}rez
et al.\,(\cite{SGLM06}), as well as for one object 
(\object{CD-23 5180}) for which
we have added a reference related to its binarity.
Note, in addition, that there are seven objects classified in SIMBAD as
{\it RV Tau} (\object{V1899\,Sgr}, \object{BD+26\,2763}, \object{MT\,Lyr},
\object{GLMP\,255}, \object{V411\,Sco}, \object{V400\,Sco}, 
\object{V558\,Sgr}),
with no bibliographic entries in ADS (SIMBAD). These 7 RV Tauri
sources do not have links to ADS in the bibliography field of our catalogue.

The main bar above the list of objects contains a number of buttons:

\begin{itemize}

\item {\bf Home} -- allows the user to get back to the entry screen;

\item {\bf Change catalogue} -- allows the selection of one of the
sub-catalogues;

\item {\bf Info} -- show some additional information about the catalogue;

\item {\bf Search} -- allows the user to search the sub-catalogue using
different criteria (name, class, coordinates, bibliography, etc...);

\item {\bf Browse} -- re-opens the full list of objects contained in
the selected sub-catalogue;

\item {\bf Bibliography} -- provides the full listing of the selected
bibliography;

\item {\bf Export} -- allows the user to export the required data
(see below for details).

\end{itemize}

\begin{figure*}[!bt]
\begin{center}
\centering
\resizebox{\hsize}{!}{\includegraphics[]{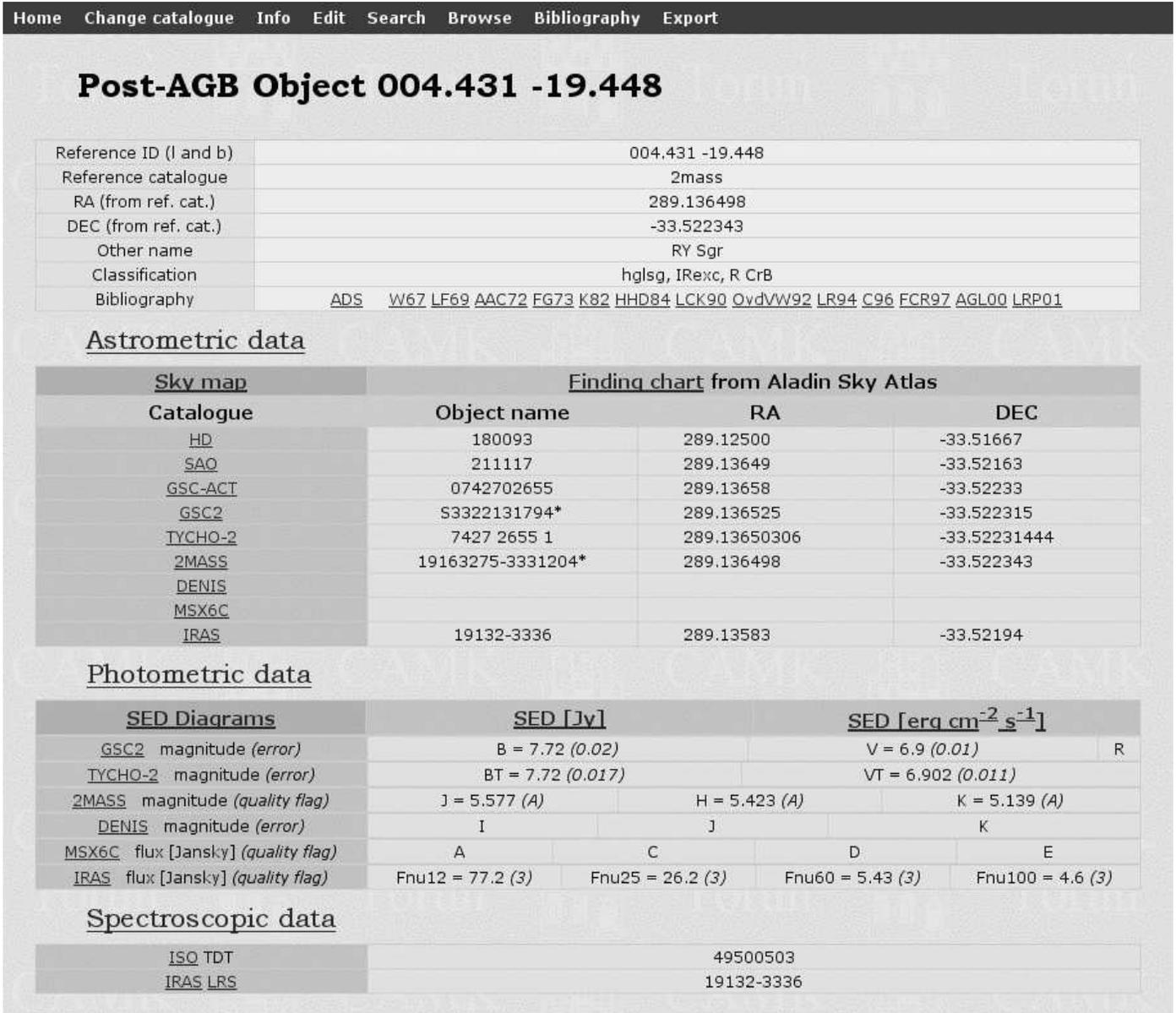}}
\caption{An example of the screen shown for a selected object.}
\label{Fig1}
\end{center}
\end{figure*}

When clicking on the $l; b$  value of a given object (this option
exists only for the table of {\it very likely}
objects) the screen as shown in Fig.\,\ref{Fig1} appears.
The main bar remains the same as on the screen described above. It
gives all the characteristics of the object available in the
catalogue:
the names\footnote{The sign * appearing after an object name
means that this counterpart was identified by us and is not shown by SIMBAD.}
and coordinates in degrees as given in the various surveys\footnote{The
description of each survey can be obtained by clicking on the
corresponding name in the screen shown in
Fig.\,\ref{Fig1}.}, the photometric data, and the spectroscopic data:
the ISO SWS\,01
spectroscopic data (reduced by one of the authors: NS) and
the IRAS Low Resolution Spectra (LRS)\footnote{The LRS spectra have
the absolute
calibration corrected as in Volk \& Cohen\,(\cite{VC89}) and Cohen, Walker \&
Witteborn\,(\cite{CWW92}).} taken from the Kevin Volk's Home Page at
the University of Calgary
http://www.iras.ucalgary.ca/$~$volk/getlrs$_{-}$plot.html.
If the user clicks on the "Finding chart" button, a finding chart is then 
displayed, generally extracted from 2MASS and from the ESO Digitized
Sky Survey-2
Red (see http://archive.eso.org/dss/dss). When clicking on the ``SED"
button, the spectral energy distribution appears in [Jy] or in
[erg\,cm$^{-2}$\,s$^{-1}$], as a
function of wavelength, for all the collected photometric data. If
ISO SWS\,01 and/or IRAS LRS spectra are
available, then they are overplotted on the SED diagram in corresponding units.

The  data from our catalogue can be exported by clicking on the
``Export" button - in the screen shown in
Fig.\,\ref{Fig1}. The desired format for the exported table is ascii or Excel.

We plan to follow the literature in order to update the catalogue on
a regular basis. The changes introduced
will be successively marked in the log file (to be opened
when the updating procedure starts). 
Authors of papers on post-AGB objects are invited to send
their comments and/or papers directly
to postagb@ncac.torun.pl. However, it will be our responsibility to
include new objects in the
catalogue, following the precepts given in Section\,2.

\section{Some statistical properties of post-AGB objects in our catalogue}

Since the objects appearing in our catalogue have been discovered by
various authors and using different
selection criteria, there is no well defined selection function.
Therefore, our catalogue is not the best
suited for population studies. Such studies require well defined
samples, e.g. as those obtained by
follow-up studies of well defined candidates (see a 
colour-selected, flux-limited sample of IRAS sources
described in Su{$\acute{\rm a}$}rez et al.\,\cite{SGLM06}). On the
other hand, our catalogue can be useful when looking
for correlations between various properties of post-AGB objects (e.g.
Stasi\'nska et al.\,\cite{SSS06} or Si\'odmiak et al.\,\cite{SMU07})
and when searching for evolutionary
connections between different classes of post-AGB objects and their
progenitors or progenies.

\begin{figure}[!bt]
\begin{center}
\centering
\resizebox{\hsize}{!}{\includegraphics[]{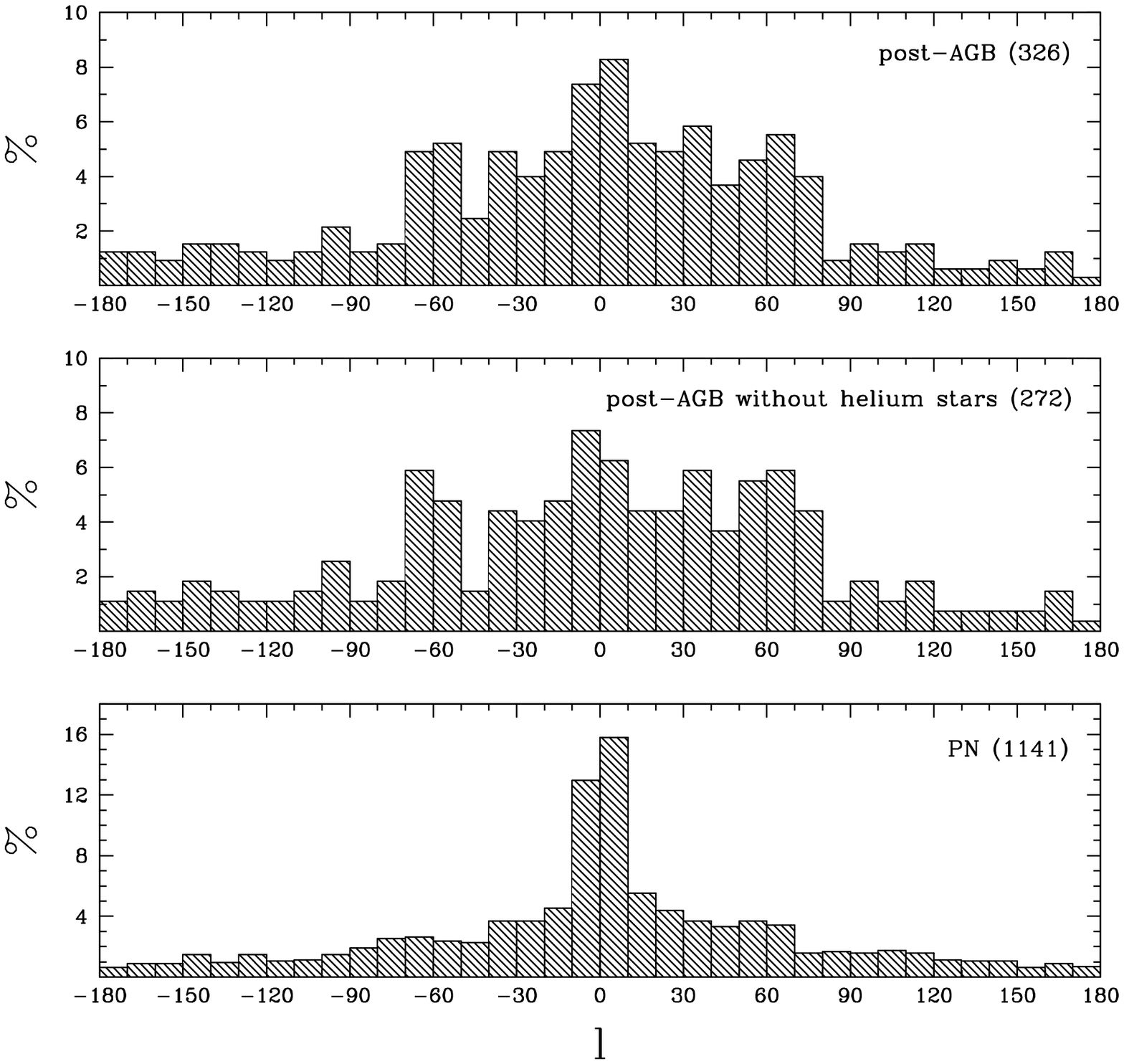}}
\caption{Galactic longitude distribution for all {\it very likely}
post-AGB objects from our catalogue
(top panel), for post-AGB objects with helium stars excluded: i.e.
without R\,CrB stars, extreme
helium stars and {\it LTP} objects (middle panel), and for planetary
nebulae from the Strasbourg catalogue - Acker
et al.\,(\cite{AMO92}) (bottom panel). The numbers in parentheses
show the number of objects used to make the diagrams.}
\label{Fig2}
\end{center}
\end{figure}
\begin{figure}[]
\begin{center}
\centering
\resizebox{\hsize}{!}{\includegraphics[]{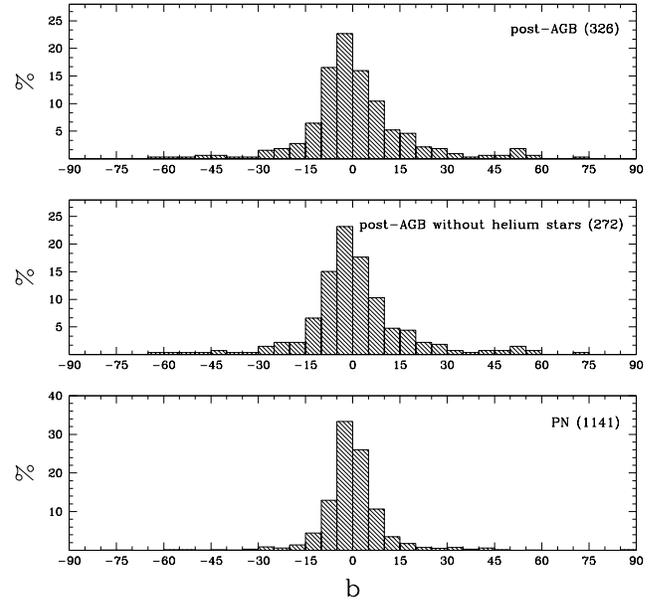}}
\caption{Galactic latitude distribution for all {\it very likely}
post-AGB objects from our catalogue
(top panel), for post-AGB objects with helium stars excluded: i.e.
without R\,CrB stars, extreme
helium stars and {\it LTP} objects (middle panel), and for planetary
nebulae from the Strasbourg catalogue - Acker
et al.\,(\cite{AMO92}) (bottom panel). The numbers in parentheses
show the number of objects used to make the diagrams.}
\label{Fig3}
\end{center}
\end{figure}

In Fig.\,\ref{Fig2} we plot the Galactic longitude distribution for
all 326 {\it very likely} post-AGB objects
(top panel) and for the sample of 272 post-AGB objects with helium
stars excluded: i.e. without
R\,CrB stars, extreme helium stars and {\it LTP} objects (middle
panel). Both panels show a similar distribution
with a concentration of objects towards the Galactic center and a
significant proportion of objects in the range
$-$60\degr $\le$ $l$ $\le$ $+$60\degr. The similarity of both
diagrams means that the
distribution of helium stars, in fact, does not differ from that of
the remaining post-AGB objects. The bottom panel
in this figure shows the Galactic longitude distribution for
planetary nebulae taken from the Strasbourg catalogue
(Acker et al.\,\cite{AMO92}). The distributions of planetary nebulae
and post-AGB objects are clearly different.
One reason is that planetary nebulae detected from strong emission lines are
easily spotted, even in the Galactic bulge, while
this is not possible, in general, for post-AGB objects. On the other
hand, low-mass, post-AGB stars will not be able
to develop planetary nebulae so there may be more lower-mass objects
among post-AGB sources. This is seen
in Fig.\,\ref{Fig3} (the structure of this figure is same as that of
Fig.\,\ref{Fig2}), which compares Galactic latitude
distributions for post-AGB objects and planetary nebulae.
There is a far smaller proportion of ``high Galactic
latitude'' planetary-nebulae than {\it hglsg} post-AGB
objects, which, statistically speaking, 
are probably less massive. However,
since post-AGB sources are statistically closer than planetary
nebulae, the proportion of ``high Galactic latitude'' objects
may be relatively larger in the sample of ``closer'' objects.
Note that y-scales are the same for the all panels of each figure.

\begin{figure}[!b]
\begin{center}
\centering
\resizebox{\hsize}{!}{\includegraphics[]{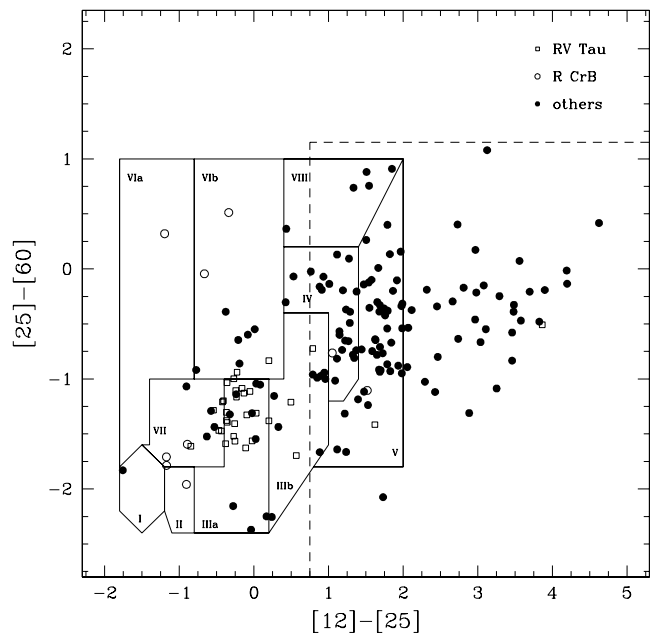}}
\caption{IRAS colour-colour diagram for 176 {\it very likely} post-AGB
objects with IRAS flux quality {\it Q}=3 or 2
at all 3 bands: 12, 25 and 60\,$\mu$m. RV Tauri stars are marked by
open squares, R CrB by open circles,
while other objects are represented by filled circles. 
The regions defined by van der Veen \&
Habing\,(\cite{vdVH88}) are shown by solid lines, while dashed ones demarcate
a typical region of search for post-AGB objects -- see text for details.}
\label{Fig4}
\end{center}
\end{figure}

As mentioned before, the  IRAS colour-colour diagram was frequently
used to search for post-AGB candidates
(e.g. Kwok et al.\,\cite{KBH87}; van der Veen et al.\,\cite{vdVHG89};
Hu et al.\,\cite{HSdJ93},
Garcia-Lario\,\cite{GLMP97}  or, most recently, Su{$\acute{\rm
a}$}rez et al.\,\cite{SGLM06}). The
largest group among {\it very likely} post-AGB objects is, in fact, the
group of {\it IRASsel} sources (115). In
Fig.\,\ref{Fig4} we present IRAS
[25]-[60]\,=\,2.5\,log[F$_{60}$/F$_{25}$] versus
[12]-[25]\,=\,2.5\,log[F$_{25}$/F$_{12}$] diagram for all 176 {\it
very likely} post-AGB objects with IRAS flux
quality, {\it Q}, equal to 2 or 3.
The zones defined by van der Veen \& Habing\,(\cite{vdVH88}) are 
shown for reference. A
typical region of search for post-AGB objects with {\it colors
like planetary nebulae} is marked by dashed lines (see also Fig.\,3 
in Su{$\acute{\rm
a}$}rez et al.\,\cite{SGLM06}) and is defined by [12]-[25] $>$ 0.75 
and [25]-[60]
$<$ 1.15 . In  Fig.\,\ref{Fig4} there are 59 {\it very likely}
post-AGB objects, which are located outside this region. These sources are
predominantly RV Tauri and R CrB stars, but not exclusively.  It is clear
that colour-colour selection, while important, cannot be the only way of
selecting post-AGB candidates.

\begin{figure}[!bt]
\begin{center}
\centering
\resizebox{\hsize}{!}{\includegraphics[]{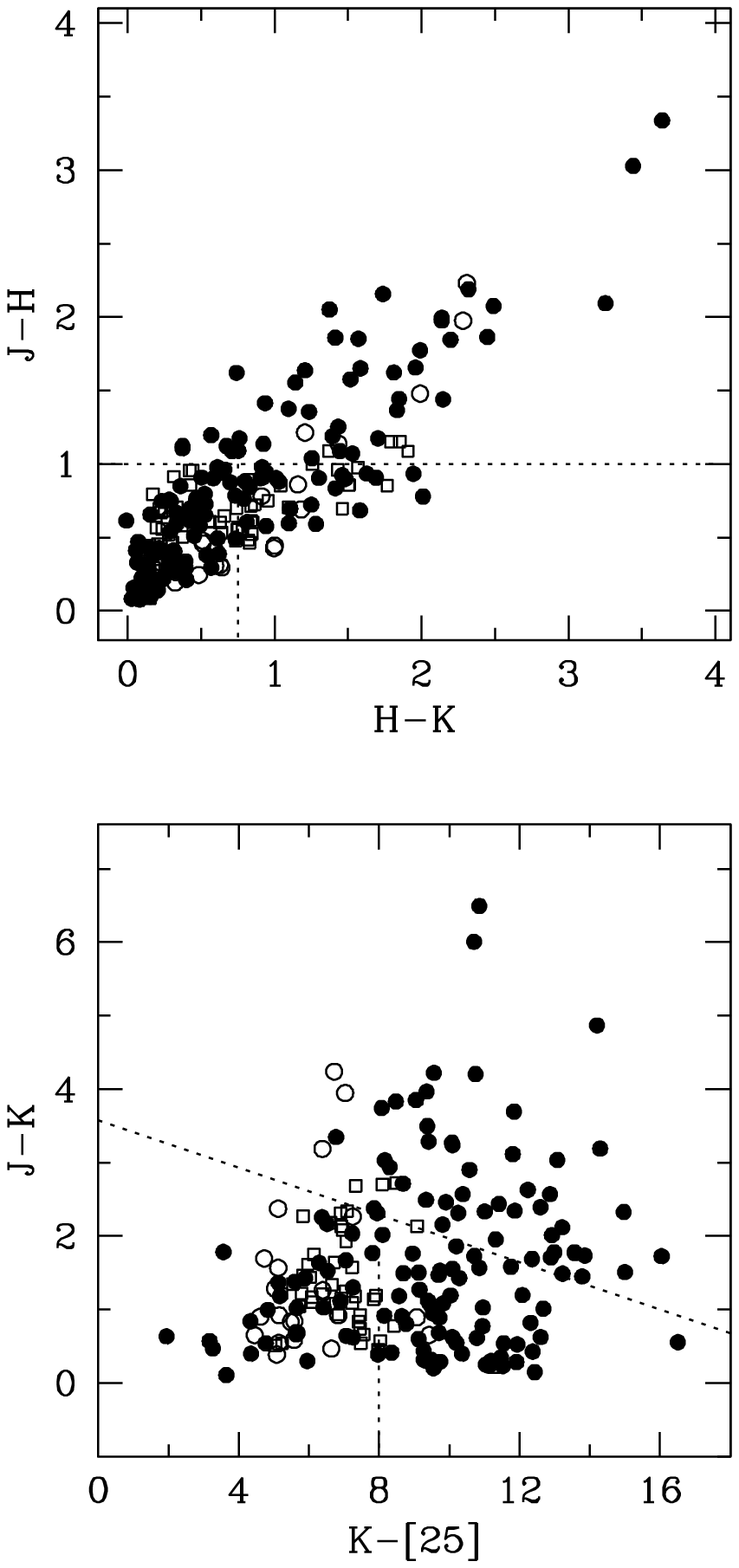}}
\caption{Infrared color-color diagrams for {\it very likely} post-AGB
objects from our catalogue. RV Tauri
stars are marked by open squares, R CrB by opens circles, while other
objects by filled circles. Dashed
lines separate parts of the diagrams occupied by different group of
objects (see Si\'odmiak et al.\,\cite{SMU07}).}
\label{Fig5}
\end{center}
\end{figure}
The photometric data that we collected from IRAS and 2MASS catalogues
allow us to construct and analyze color-color
diagrams based on these data. Fig.\,\ref{Fig5} shows J$-$H versus
H$-$K (top panel) and J$-$K versus
K$-$[25] (bottom panel) diagrams for {\it very likely} post-AGB objects
from our catalogue. The fluxes through the
    2MASS filters were transformed to the Johnson system
(Carpenter\,\cite{C01}) and the IRAS flux at
25\,$\mu$m was converted to magnitude using
[25]=-2.5\,log(F$_{25}$/6.73) (Beichman et al.\,\cite{BNH88}).
The dotted lines in Fig.\,\ref{Fig5} separate different groups of
objects and are
introduced by Si\'odmiak et al.\,(\cite{SMU07}). DUPLEX sources (as
defined by Ueta et al.\,\cite{UMB00})
have optically thick circumstellar envelopes (they have, therefore, the
most red J$-$H and J$-$K colors) and
are located above the dashed lines (horizontal and
inclined, for the top and bottom panel, respectively) in both
diagrams. SOLE objects (see Ueta et
al.\,\cite{UMB00}) are located below these lines and have
H$-$K$<$0.75 (top panel) and K$-$[25]$>$8
(bottom panel) as expected for sources with clearly visible
central stars and thin dust envelopes.
The remaining objects have H$-$K$>$0.75 (top panel) and K$-$[25]$<$8
(bottom panel) and are called
stellar (no signature of circumstellar envelope in the Hubble Space
Telescope images) by  Si\'odmiak et
al.\,(\cite{SMU07}). As can be seen, most of RV Tauri (open squares)
and R CrB stars (open circles)
are stellar or SOLE objects, in agreement with their relatively small
infrared excess.

\begin{figure}[!bt]
\begin{center}
\centering
\resizebox{\hsize}{!}{\includegraphics[]{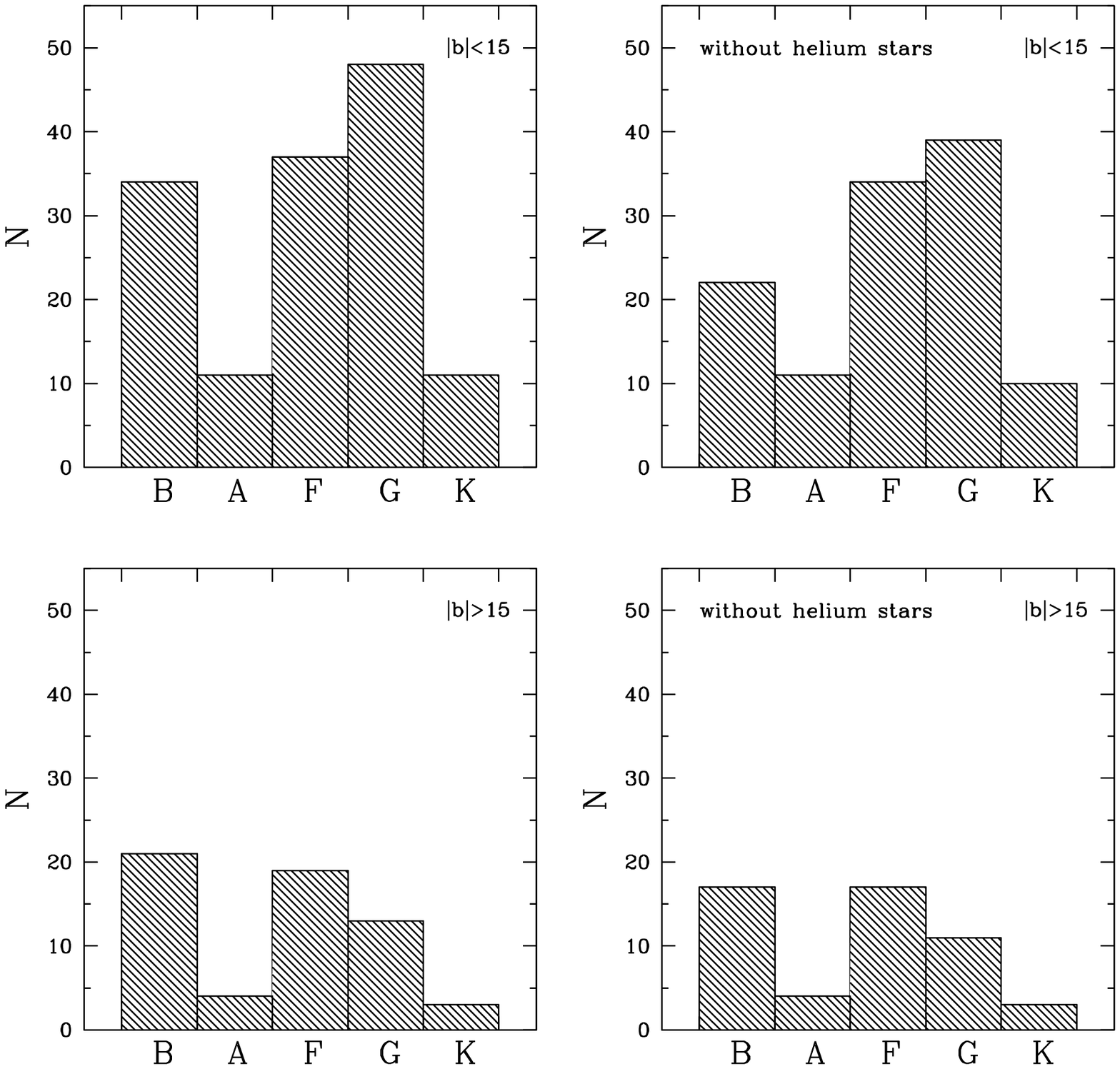}}
\caption{Distribution of spectral types for {\it very likely} post-AGB
objects at low (top panels) and high
(bottom panels) Galactic latitudes.  The right panels show the
distribution for post-AGB objects with helium stars
excluded:, i.e. without R CrB stars, helium stars and LTP objects.}
\label{Fig6}
\end{center}
\end{figure}
Finally, Fig.\,\ref{Fig6} shows four panels with the distribution of
spectral types for the {\it very likely}
post-AGB objects in our catalogue. The top panels correspond to
post-AGB objects located
relatively close to the Galactic plane, while the  bottom ones
correspond to  high Galactic latitudes ($|b|>$15{\degr}) sources. The right
panels show the spectral
type distributions after removing helium stars from the sample.

The most striking feature seen in all four panels is the gap around
10000 K (spectral type A). The
existence of such a gap can be inferred from theoretical models (see
e.g. Fig. 6 in Bl\"ocker\,\cite{B95}) and
is related to the evolutionary rates during the post-AGB phase of
stellar evolution. A similar gap was already observed
in the smaller sample of about 220 post-AGB objects discussed by
Szczerba et al.\,(\cite{SGZJ01}) and is also
present in the distribution of spectral types for the
colour-selected sample of
Su{$\acute{\rm a}$}rez et al.\,(\cite{SGLM06}).As with the
Galactic distribution of post-AGB
objects discussed above we do not see any significant influence of
helium post-AGB stars on the observed
distribution of spectral types. This may suggest that the spectral
evolution of H-rich and He-rich objects
during the post-AGB phase is similar as far as timescales are
concerned. On the other hand, it seems that there
is a difference in the distribution of spectral types for {\it hglsg}
and for sources lying closer to the
Galactic plane. This is so because there are more F-type than G-type sources
among {\it hglsg} objects, while the opposite
is true for post-AGB objects at lower-Galactic latitudes. This
finding seems to contradict 
theoretically predicted timescales of evolution, i.e. featuring
mass loss during the post-AGB phase
(Trams et al.\,\cite{TWW89}). We expect that there are  more massive
objects among those
with lower-Galactic latitudes; however, the massive tracks of
Bl{\"o}cker\,(\cite{B95}) predict more F-type
stars than G-type ones (see e.g. Fig.2 in van Hoof et
al.\,\cite{vHOW97}). Definitive statements about timescales during the
post-AGB phase, based on the statistics of post-AGB objects, require, 
however, a
more complete  sample than the one presented in this catalogue together with a
proper correction for selection effects.

\section{Summary}

By searching the available literature we were able to identify 326
{\it very likely} post-AGB objects in our
Galaxy. The collected information was organized into an evolutionary,
on-line catalogue of
Galactic post-AGB objects, hereafter called {\it the Toru\'n catalogue
of Galactic post-AGB and related objects}. The present version
contains 326 {\it very likely},
107 {\it possible} and 64 {\it disqualified} post-AGB objects. It is
anticipated that the number of known post-AGB objects will increase
significantly in the near future, owing to systematic searches based on
infrared surveys. We plan to regularly update our  catalogue after
new discoveries.   The catalogue can be useful when looking for
correlations between various properties of post-AGB objects and 
searching for evolutionary
connections between different classes of post-AGB objects and their
progenitors or progenies.
The catalogue has already been successfully used in the work by
Stasi\'nska et al.\,(\cite{SSS06}), which showed
that post-AGB stars, as a class, are valuable testbeds for theories
of AGB nucleosynthesis and in the work by
Si\'odmiak et al.\,(\cite{SMU07}) devoted to the Hubble Space Telescope
snapshot survey of post-AGB objects
characterized by different physical and chemical properties.

\begin{acknowledgements}
This work has been partly supported by grant 2.PN203 019 31/2874
awarded by the Polish State Committee for Scientific Research and by the
European Associated Laboratory
``Astronomy Poland-France". We have made extensive use of NASA's
Astrophysics Data System Bibliographic Services and SIMBAD database.
\end{acknowledgements}

\end{document}